\documentclass[conference]{IEEEtran}

\usepackage{amsmath} 
\usepackage{amsfonts} 
\usepackage{amssymb} 
\usepackage{booktabs}
\usepackage{bm} 
\usepackage{lipsum} 
\usepackage{hyperref}
\usepackage{pgfplots}

\usepackage{todonotes}
\usepackage{graphicx} 
\usepackage{tablefootnote}
\usepackage{todonotes}

\usepackage{listings}

\definecolor{codebackground}{rgb}{0.9,0.9,0.9}
\definecolor{pythonkeyword}{RGB}{177,92,171}
\lstdefinestyle{pythonExamples}{
    breaklines=true, 
    numbers=left, 
    numbersep=5pt,
    numberstyle=\tiny,
    backgroundcolor=\color{codebackground},
}
\usepackage[style=ieee,citestyle=numeric-comp,backend=biber]{biblatex}
\DeclareRobustCommand{\IEEEauthorrefmark}[1]{\smash{\textsuperscript{\footnotesize #1}}}

\title{Learn to Code Sustainably: An Empirical Study on LLM-based Green Code Generation} 

\date{}

\author{
    \IEEEauthorblockN{ Tina Vartziotis\IEEEauthorrefmark{1}\IEEEauthorrefmark{2}, Ippolyti Dellatolas\IEEEauthorrefmark{3}, George Dasoulas\IEEEauthorrefmark{4}, Maximilian Schmidt\IEEEauthorrefmark{1}, \\  Florian Schneider\IEEEauthorrefmark{1}, Tim Hoffmann\IEEEauthorrefmark{5}, Sotirios Kotsopoulos\IEEEauthorrefmark{2}\IEEEauthorrefmark{3}, Michael Keckeisen\IEEEauthorrefmark{1}}%
    
    \IEEEauthorblockA{\IEEEauthorrefmark{1}TWT GmbH Science \& Innovation}
    \IEEEauthorblockA{\IEEEauthorrefmark{2}National Technical University of Athens}
    \IEEEauthorblockA{\IEEEauthorrefmark{3}Massachusetts Institute of Technology}
    \IEEEauthorblockA{\IEEEauthorrefmark{4}Harvard University}
    \IEEEauthorblockA{\IEEEauthorrefmark{5}Mercedes-Benz}
}

\usepackage{amsmath,amsfonts,bm}

\def\eqref#1{equation~\ref{#1}}

\def\1{\bm{1}}

\def\vc{{\bm{c}}}

\def\mX{{\bm{X}}}

\DeclareMathAlphabet{\mathsfit}{\encodingdefault}{\sfdefault}{m}{sl}
\SetMathAlphabet{\mathsfit}{bold}{\encodingdefault}{\sfdefault}{bx}{n}

\begin{document}

\maketitle

\begin{abstract}
The increasing use of information technology has led to a significant share of energy consumption and carbon emissions from data centers. These contributions are expected to rise with the growing demand for big data analytics, increasing digitization, and the development of large artificial intelligence (AI) models. The need to address the environmental impact of software development has led to increased interest in green (sustainable) coding and claims that the use of AI models can lead to energy efficiency gains.
Here, we provide an empirical study on green code and an overview of green coding practices, as well as metrics used to quantify the sustainability awareness of AI models. In this framework, we evaluate the sustainability of auto-generated code. The auto-generate codes considered in this study are produced by generative commercial AI language models, GitHub Copilot, OpenAI ChatGPT-3, and Amazon CodeWhisperer. Within our methodology, in order to quantify the sustainability awareness of these AI models, we propose a definition of the code’s “green capacity”, based on certain sustainability metrics. We compare the performance and green capacity of human-generated code and code generated by the three AI language models in response to easy-to-hard problem statements. Our findings shed light on the current capacity of AI models to contribute to sustainable software development. 

\end{abstract}
\section{Introduction}

\subsection{Environmental Impact of AI}
In the era of rapid technological advancement, the environmental impact of software development has become a pressing concern.
Advances in software and hardware techniques have led to a significant increase in the demand for digital services, with internet traffic increasing 25-fold since 2010 \cite{iea2023}. Global data center electricity consumption, excluding cryptocurrency mining, was estimated to be 240-340 TWh in 2022, representing 1-1.3\% of global electricity consumption and 0.3\% of gloal carbon emissions \cite{iea2023, masanet2020recalibrating, hintemann2020}. Countries with expanding data center markets are seeing rapid data center electricity consumption growth. For example, data center electricity consumption in Ireland accounted for 20\% of the country's total electricity use in 2022, triple its 2015 levels \cite{IrelandCSO2023}. Energy efficiency gains have helped limit data center energy demand, including improvements in IT hardware, cooling systems, and a shift from local processors to large-scale datacenters; yet increased computing demand for applications such as blockchains and machine learning could outpace efficiency gains.
Artificial intelligence and machine learning are areas where the demand is expected to grow. A 2018 OpenAI study found that the amount of compute (in $10^{15}$ floating point operations per second - day) has been increasing exponentially since 2012 due to algorithmic innovation and the development of new AI models \cite{openai2018}. 
The lifecycle emissions of an AI model encompass the emissions generated during the development of the model (training and testing phase), as well as those resulting from running the model once it is deployed (inference phase or queries). 
Several studies have estimated the training cost of AI models, including natural language processing (NLP) models that are highly computationally expensive \cite{strubell-etal-2019-energy, Patterson2021, Schwartz2020}. 
Estimated $\mathrm{CO}_2$ emissions from training a large NLP model are 1-10 times those from the lifecycle of a car \cite{strubell-etal-2019-energy}.
The energy consumption of large NLP models during training ranged from 20MWh to more than 1200MWh, with the development of GPT-3 estimated to have generated 552 tons of $\mathrm{CO}_2$ equivalent ($\mathrm{tCO}_2e$) \cite{Patterson2021}. These emissions only encompass the training phase of the models. Once the models are deployed, their usage triggers additional energy demand due to consumer queries. The energy consumption of the most cutting-edge computer vision and NLP models during inference was found to have increased exponentially \cite{desislavov2023}, with total emissions depending on the number of consumer queries, or how frequently the model is used (known as the multiplicative factor).
Recent generative AI language models have already experienced increased demand, triggering large numbers of queries since their release, amounting to an energy consumption in the inference phase that is a significant contributor to the model lifecycle emissions \cite{wu2022}. For example, assuming 100 million daily queries - estimated from data from SimilarWeb - and an energy use per query of roughly 0.002kWh - based on the average energy required by a forward pass of an NLP model \cite{desislavov2023} - the energy consumption of ChatGPT can be estimated to be about 0.2GWh per day.

Despite the significant carbon footprint associated with the development and use (training and inference) of AI models, they are said to have the potential to contribute to environmental sustainability. This potential is encapsulated in the "carbon handprint" of software, where the use of the software leads to energy savings \cite{Schwartz2020}. One application of generative AI models is code generation. In this paper, we evaluate the readiness of generative AI language models to produce sustainable code. This type of code combines the ability to successfully perform a task with optimizing sustainability metrics, leading to lower energy consumption and carbon emissions. 

The need for such green coding practices is particularly pressing in the current global context, where adopting heat pumps, electric vehicles, and electrification will generally increase the demand for clean energy \cite{iea2022}. Thus far, data center energy demand has outpaced the deployment of renewable energy \cite{acun2023}. This gap underscores the importance of developing and using energy-efficient code, thus reducing the overall environmental impact of digital infrastructures. Our study aims to assess the capability of generative AI models to generate code that aligns with these sustainability goals, considering the ongoing global shift towards more renewable energy sources. By doing so, we hope to provide insights into how these advanced technologies can be harnessed to support environmental sustainability in the realm of software development.

\vspace{-.3cm}

\subsection{Scope of This Paper}
This paper is motivated by the urgent need to address sustainability challenges in software engineering. 
By introducing metrics to gauge software "greenness," we aim to shed light on the significance of green coding. We explore whether the sustainability of code can be enhanced through the utilization of automated code-generation tools. There are various methods to automatically generate code, including \textit{enumerative search} \cite{PGL-010, DBLP:conf/pldi/HuangQSW20}, \textit{stochastic search} \cite{ellis2019write}, or \textit{constraint solving} \cite{10.1145/1836089.1836091}, as well as AI-based methods utilizing, for example, natural language processing. In this paper, we specifically focus on the AI-based tools \href{https://chat.openai.com/}{OpenAI ChatGPT}, \href{https://github.com/features/copilot}{GitHub Copilot}, and \href{https://aws.amazon.com/codewhisperer/}{Amazon CodeWhisperer}. We investigate the potential benefits and challenges associated with leveraging these tools to streamline the software development process, improve code quality, and reduce the environmental impact of source code. Additionally, the study will delve into the effectiveness of these tools to generate code optimized for sustainability, including optimizations of resource utilization. The paper aims to provide insights into the practical implementation of automated code generation tools and their contribution to sustainable software development, evaluating their potential to drive positive environmental and efficiency outcomes.

To this end, we compare the sustainability metrics of code written by humans, the initial, and the optimized output of the AI-generated code. We define \textit{Green Capacity} (GC) as the degree of eco-friendliness of the model based on the specific problems, gauged against selected sustainability metrics. It encapsulates the code's correctness, as well as its adherence to principles of energy efficiency, and its environmental impact, providing a holistic measure of its sustainability profile.

We assess the performance of automated code generation tools using six programming problems and analyze their sustainability regarding various parameters significantly impacting green coding. The programming problems are chosen from \href{https://leetcode.com/}{LeetCode}~\cite{leetcode}, a platform that trains software engineers to develop efficient algorithms for problem-solving. The same tasks, which are popular problems, are being requested by the three language models to generate a solution. These are not long programming tasks or solutions, but the focus is to find problems that have been widely used and ensure that the AI can find correct solutions and potentially optimize them based on the sustainability metrics as requested.
We compare the methods in terms of successful code generation and green capacity. Our study identifies which method generates, on average, faster/slower code, which generated code consumes more/less energy, and memory usage. Finally, we explain how our findings can potentially assist software engineers in choosing an appropriate method when sustainability is a concern. There is a continuous evolution in the automated code generation tools, such as new versions or new tools being developed, such as \href{https://www.ibm.com/products/watsonx-code-assistant}{IBM Watson X Code Assistant}, and \href{https://chat.openai.com/}{OpenAI ChatGPT 4.0}, which are not considered in the framework but should be included in future extensions of this analysis.
Our evaluation code is available \href{https://anonymous.4open.science/r/green-coding-review-BDB3/}{here}.

\vspace{-.2cm}
\section{Sustainability Metrics and Green Capacity}

\subsection{Green Coding Practices}

We first review how green coding is defined in the literature and clarify which aspects of green coding we tackled here. The term green coding is generally used to describe a practice that reduces software's final carbon footprint. Given that the final carbon footprint (in $\mathrm{tCO}_2e$) of the code is the product of its energy use (in kWh) and of the carbon intensity of the energy required to run it (in $\mathrm{tCO}_2e$/kWh), emissions reductions can be achieved by either making the software itself more efficient or by relying on less carbon-intensive energy. Purchasing carbon offsets, for example, could lead to lower reported emissions for software, without any gains in its energy efficiency. Within this broad landscape, our study focuses on the energy efficiency of the software itself. For AI models, the energy used by the software can be decomposed into the energy usage to run the model operations and the energy required to cool the data center in which the code is run \cite{Patterson2021}. The energy used to run the model operations itself depends on the energy efficiency of the hardware used, as well as on the structure of the software. 
Most previous work has focused on reducing the carbon footprint of AI models during the training phase, with some studies targeting the inference phase \cite{Schwartz2020}.
Practices aiming to reduce the carbon footprint of software include:

\begin{itemize}
    \item reducing the carbon emissions associated with electricity consumption, by placing the data center in a location where the grid is cleaner, or purchasing low-carbon electricity to power the data center \cite{Henderson2020}
    \item improving the energy efficiency of the data center by reducing its power usage effectiveness (PUE, the ratio between the total data center energy usage and the energy required to power computing equipment): this generally translates to reducing the energy required to cool the data center, through infrastructure optimization or by locating it in a colder climate \cite{Lazic2018}
    \item improving the hardware efficiency by using processors that require less power for a given performance (FLOPs) \cite{Lacoste2019}
    \item improving the efficiency of the algorithm, for example, through changes in the code architecture \cite{So2019}
\end{itemize}

This study does not account for variability in grid carbon intensity, PUE, or hardware. Indeed, our evaluation test cases are all run on the same hardware and grid. We, therefore, compare codes where the only gain in efficiency can be attributed to changes in the generated code itself, like the algorithmic structure, using more efficient or more suitable instructions.

\subsection{Sustainability Metrics}


Machine learning research papers rely on different metrics to quantify how computationally expensive and sustainable a code is. These include the runtime \cite{Jeon2018, Qin2018, Patterson2021}, the memory usage \cite{Henderson2020}, the number of floating point operations (FLOPs) \cite{David2010, Sandler2018, Schwartz2020, desislavov2023, Patterson2021}, the number of parameters in the neural network \cite{David2010, Sandler2018}, the percentage of GPU or CPU utilization \cite{Assran2019, Dalton2020}, the energy consumption in Joules or MWh \cite{Assran2019, Patterson2021}, the power consumption in Watts \cite{Canziani, strubell-etal-2019-energy}, and the CO2 emissions \cite{strubell-etal-2019-energy, Patterson2021}. 
Although many of these parameters are inter-related, their link is quite complex. For example, FLOPs are not directly proportional to the number of parameters of a model \cite{Tan2019}. Memory usage affects runtime while contributing to the code’s energy consumption \cite{Ma2018}. Runtime and energy consumption are not directly proportional but are affected by the structure and size of the model, as well as the hardware and memory usage \cite{strubell-etal-2019-energy}. Moreover, it has been argued that FLOPs are the metric most representative of a code’s sustainability, as it is agnostic of hardware and can be computed analytically \cite{Schwartz2020}. However, the link between FLOPs and energy consumption or runtime is not always straightforward due to dependence on memory usage, firmware, and hardware usage \cite{David2010, Jeon2018, Qin2018}. It was further demonstrated that FLOPs correlate with energy consumption and runtime for a given model architecture, but the correlation does not persist across model architectures \cite{Henderson2020}. Therefore, while FLOPs can be a good measure of how sustainable a code is, they cannot be used as a direct quantification of energy consumption or runtime. 
We thus report runtime, memory usage, FLOPs, and energy consumption in addition to code correctness.

Details of the metrics are as follows:

\begin{enumerate}
    \item \textbf{Code Correctness}: Evaluate whether the generated code is correct. We consider correctness a sustainability metric because incorrect code requires additional effort to be corrected, making it unsustainable. It is a binary value, with \textit{0 indicating incorrect code} and \textit{1 indicating correct code}.
    \item \textbf{Runtime}: The time the code takes to execute and produce results. It is an essential metric as longer runtimes can indicate inefficiencies in the code. Measured in \textit{seconds (sec)}.
    \item \textbf{Memory}: The peak amount of memory space the code utilizes during its execution. Efficient memory usage is crucial for sustainability, especially in systems with limited resources. Measured in \textit{Kibibytes (KiB)}.
    \item \textbf{FLOPs}: Floating-Point Operations, representing the total number of floating-point operations that were performed during the execution of the code. It provides insights into how optimized an algorithm is by reducing the expensive computations required to solve a certain problem. Measured in \textit{number of FLOPs}.

    Note that two commonly used and similar acronyms in the literature are related to floating point operations \cite{DESISLAVOV2023100857}. One version (usually written FLOPS) describes the floating-point operations per second, which is a hardware performance metric, whereas the other (usually written FLOPs) is the amount of floating-point operations executed to perform some computation which is hardware-agnostic. In our paper, we solely focus on the latter version, namely the total number of floating-point operations performed during the execution of code.
    \item \textbf{Energy Consumption}: The amount of energy the code consumes during its execution. Lower energy consumption is indicative of a more sustainable and efficient code. Measured in \textit{Joules (J)}. 
\end{enumerate}

These metrics provide a comprehensive framework to evaluate the sustainability of the generated code in our experiments.

\subsection{Green Capacity}\label{sec:green_capacity}
Let us denote by $\mathcal{M}$ the code generation method, that takes as inputs: i) the examined problem definition prompt $\mathcal{T}$, and ii) the optimization request prompt by $\mathcal{P}=\{P_\text{init}, P_r, P_m, P_e, P_f\}$, where $P_\text{init}$ is the request for code generation on the given task, without further optimization, and $P_r, P_m, P_e, P_f$ are the requests for code generation, with respect to runtime, memory usage, energy consumption, and FLOPs minimization, respectively.
Let us, also, denote as $\mX_{s}^{(p)}$ the recorded value of the sustainability metric $s \in $ \{runtime, memory, energy consumption, FLOPs count\}
have the set $\mathcal{E}$ for an executed code generation model $\mathcal{M}$ with a given optimization prompt $p \in \mathcal{P}$. Moreover, we define a correctness indicator variable $\vc^{(p)} \in \{0,1\}$, that denotes the validity of the produced code from a generation model ($1$ in the case of valid code, $0$ otherwise). Finally, we introduce the \textit{Filtered Performance Delta} (PD) function as follows:
\begin{equation}\label{eq:pd}
    \text{PD}(a, b, c) = (\frac{a - b}{a}) * c,
\end{equation}
that we will essentially use as a filtered percentage difference between the initial and the optimized values. The filter masks the PD values, based on whether the optimized code is valid. Given the previous notations, we can now define \textit{Green Capacity} (GC) of a code to quantify its degree of eco-friendliness based on the selected sustainability metrics as a function of the code generation method and the examined task. Specifically, GC reads as:
\begin{equation} \label{eq:green_capacity}
    \text{GC}(\mathcal{M}, \mathcal{T}) = \sum_{i \in \mathcal{P}} \max\Bigg( \texttt{PD}\Big( X_i^{(P_\text{init})}, X_{i}^{(i)}, \vc^{(i)}\Big), 0 \Bigg).
\end{equation}

We note that for the computation of GC we consider the terms $X_i^{(i)}$, which means we only record the same evaluation metric with the given optimization request.
Moreover, Equation~\ref{eq:green_capacity}  shows that the need for utilizing normalized differences is that each criterion in $\mathcal{E}$ is computed in different scales, and their unnormalized combination would impose a great bias. The clamping applied in Equation~\ref{eq:pd} is motivated by our decision to consider any deterioration of the sustainability metrics as a null improvement. 

Similarly to the definition of GC with respect to the improvement of the optimized code to the initial implementation, we can define GC with respect to the human implementation (retrieved from community contributions):
\begin{equation} \label{eq:green_capacity_human}
    \text{GC}(\mathcal{M}, \mathcal{T}) = \sum_{i \in \mathcal{P}} \max \Bigg( \texttt{PD}\Big( X_i^{(P_\text{init})}, X_{i}^{\textbf{*}}, \vc^{\textbf{*}}\Big), 0 \Bigg),
\end{equation}
where $X_{i}^{\textbf{*}}$ is the $i$-th reported sustainability metric for the human implementation, and $\vc^{\textbf{*}}$ is the correctness of the human submission (it should always be equal to $1$).
For clarity reasons, we rename the GC definition in Equation~\ref{eq:green_capacity} as $\text{GC}_\text{AI}$, and the definition in Equation~\ref{eq:green_capacity_human} as $\text{GC}_\text{human}$. In these formulations, \(GC_{\text{AI}}\) quantifies the summation of relative improvements across the four performance dimensions brought forth by the AI optimization. On the other hand, \(GC_{\text{human}}\) yields the effectiveness of human-authored code relative to the initial AI-generated code across the same performance dimensions.

\section{Methodology}

\begin{figure}[ht]
    \centering
    \includegraphics[width=1\linewidth]{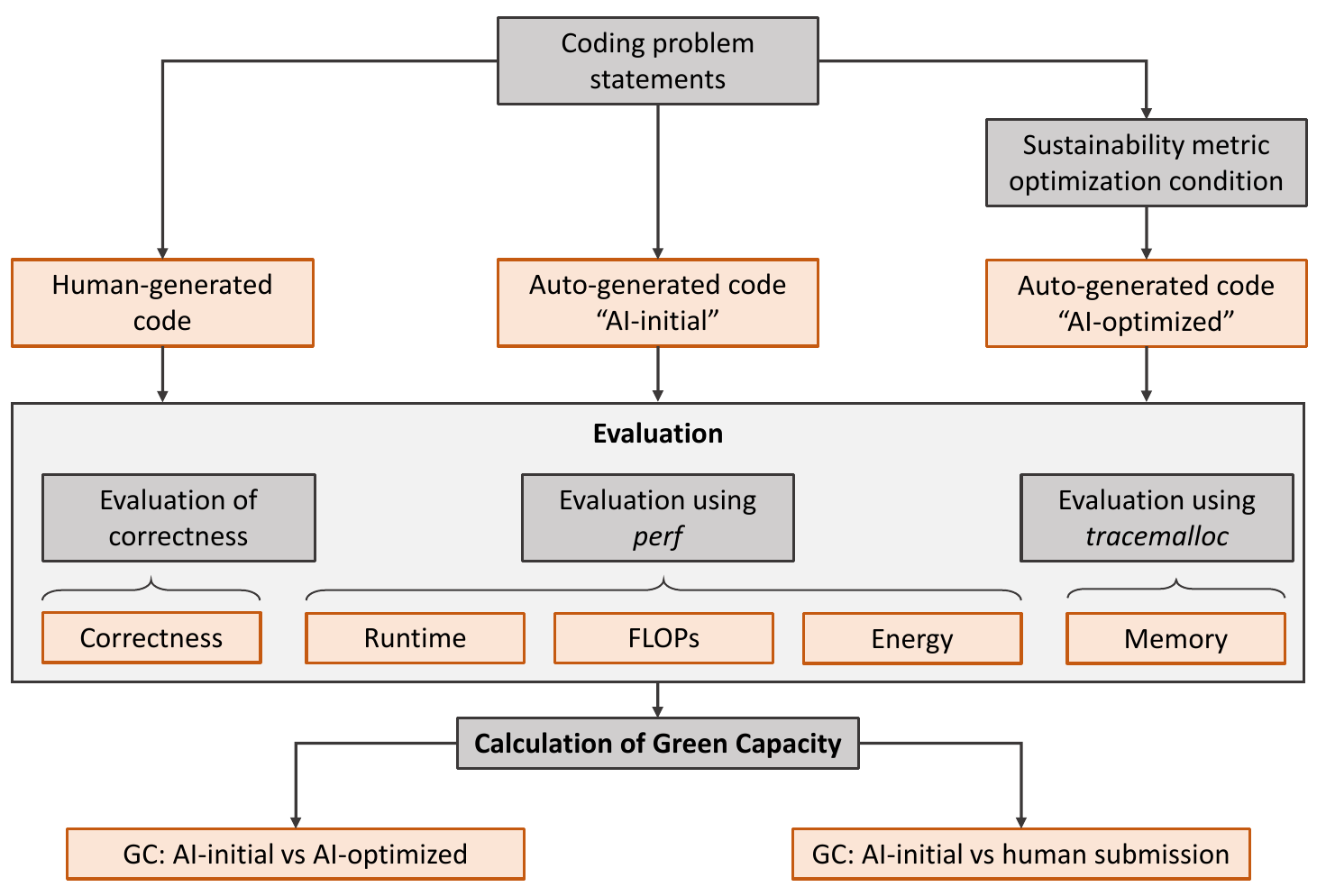}
    \caption{Evaluation steps performed for Green Capacity.}
    \label{fig:flowchart_methodology}
\end{figure}

This section outlines our methodology for assessing the Green Capacity of various tools. 
We first discuss the coding problems we selected in response to which code was generated. We further describe the process used to generate the code and detail our prompting method. Finally, we introduce the two monitoring programs employed to assess the various sustainability metrics, along with a detailed account of each program's evaluation process. A summary of the outlined procedures is depicted in Figure \ref{fig:flowchart_methodology}.

\subsection{Selection of Coding Problems}

To conduct our initial assessment, we required a set of coding tasks from which we could generate code.
At the initial stage of our process, we opted to utilize the educational platform LeetCode, which offers users a wide range of coding problems to enhance their coding proficiency.
LeetCode's problems are categorized into various topics, such as algorithmic problems, concurrency, and database-related problems.
Each problem is classified based on its level of difficulty, specifically categorized as "easy," "medium," or "hard."
LeetCode allows users to select from various programming languages to solve tasks. Python was chosen as the initial language for assessing the environmental sustainability of the tools due to its widespread usage across various domains.

We chose six problems from the algorithms domain for our experiments, spanning various difficulty levels.
Throughout the selection process, we considered the problems that allowed us to generate suitable test cases for evaluating our tasks.
To accurately assess our sustainability metrics for a sorting problem, it is necessary to have the capability to present extensive datasets in any given sequence.
Ultimately, we chose six exercises, ensuring at least one problem was selected from each difficulty level (see Table~\ref{tab:problems}).

\begin{table*}[t]
\centering
\begin{tabular}{ll|ll|lll}
\toprule
\multicolumn{2}{c|}{\textbf{Code Generation Methods}}      & \multicolumn{2}{c|}{\textbf{Sustainability Metrics}} & \multicolumn{3}{c}{{ \textbf{Coding Problem}}} \\ \midrule
AI-based      & Non-parametric     & Functional               & Environmental             & Easy                          & Medium                       & Hard                        \\ \midrule
Copilot       & Enumerative Search & Runtime                  & Energy consumption        & Cookies                       & Network                      & Median                      \\
CodeWhisperer & Constraint Solving & Memory                   &                           & Search                        & 3Sum                         &                             \\
ChatGPT       & Stochastic  Search & FLOPs                    &                           &                               & Sort                         &                             \\
              &                    & Code Correctness         &                           &                               &                              &   \\
              \bottomrule
\end{tabular}
\vspace{0.15cm}
\caption{Categorization of evaluation components for sustainable code generation. In the left box, we show the methods for automatic generation, divided into AI-based and non-parametric ones. In the middle box, we highlight the sustainability metrics we monitor to measure the automatic code generators' capacity for sustainable computing. In the right box, we divide the problem tasks into three categories based on their hardness level, characterized by LeetCode.}
\label{tab:problems}
\end{table*}

\subsection{Code Generation}

Following the selection of the coding tasks, the subsequent step involves generating code for evaluation.
In our present assessment, we specifically examined three leading coding assistant tools accessible at the time of writing: OpenAI's ChatGPT, GitHub Copilot, and Amazon CodeWhisperer.
We provided identical information to each tool, equivalent to what a human would get from LeetCode. When using ChatGPT on a web browser, it is important to mention that we utilized a single chat window for each of the six problems. This means that we have a continuous interaction with ChatGPT for each task.
To initiate the process, we copied the task description in the designated input field and included the function header provided by LeetCode. This allowed ChatGPT to utilize it as a foundation for generating the solution.

Regarding GitHub Copilot and Amazon CodeWhisperer, we had to adopt a different approach to code generation due to the different interfaces for generating code, missing the chat option. Copilot and CodeWhisperer are designed to be integrated into the development environment, they generate and suggest code to the user based on existing code in the same file or following comments made by the user in the file. To activate code suggestions, we employ the second option, which involves creating a new Python file and including a comment that describes the task as stated in LeetCode.
Immediately following that, we include the function signature provided by LeetCode. Subsequently, the coding assistant recommends solving the task described in the preceding comment.

To obtain optimized versions, we initiate the comment by providing the instructions to optimize the solution.
To provide a runtime-optimized solution, we add the description with "Q: Give me a runtime-optimized solution for this problem," followed by the task.
A separate file is generated for each code variant (including optimization, runtime, etc.) to avoid duplicating existing code and increase the possibility of optimizations or changes.
These steps were implemented for GitHub Copilot and Amazon CodeWhisperer.

To evaluate human submissions, we select from the top 0.05\% of LeetCode submissions based on runtime. Programmers, in their training phase, submit solutions on LeetCode, creating a comprehensive pool of optimized problem solutions. From this, we choose one of the best submissions to assess the differences between top human programmers and AI code generation tools.

\subsection{Experimentation Setup}

To robustly evaluate sustainability metrics, we create several test cases (see Table~\ref{tab:problems}). When relevant, we supplement the code input variables with additional randomly generated inputs to increase the problem complexity. For instance, in the case of the \textit{Search} problem, we use random number lists as inputs.
Additionally, since an individual function call does not have a long runtime, we execute each function multiple times to reduce the impact of overheads of the Python interpreter itself, object instantiations, or imports. The number of iterations we select depends on the complexity of the solution. For example, the implementation for the \textit{Search} coding problem takes 0.6 to 3 seconds for 100,000 iterations. On the other hand, the solution for the \textit{3Sum} coding problem takes between 46 and 73 seconds for 1,000 iterations. Based on each problem's complexity, we choose a number of iterations $\in \{1000,100000\}$ to generate noticeable distinctions without unnecessarily increasing the runtime. Finally, we proceed with our assessments utilizing \textit{perf} and \textit{tracemalloc} to evaluate the execution time, FLOPs, energy consumption, and maximum memory usage. We evaluate the outputs of 3 code generation model versions, ChatGPT version 3.5, GitHub Copilot Individual Version, and Amazon CodeWhisperer Individual Tier. Our tests are performed on a desktop PC with an Intel Core i7-6700k CPU with 32GB RAM, Linux Mint 21.2 Cinnamon, and kernel version 5.17.0-79-generic. Furthermore, we use perf version 5.15.111 and Python version 3.12.0rc1.

\subsection{Analysis of Sustainability Metrics}

We examine the generated code samples, focusing on sustainability metrics. Our analysis includes code correctness, memomory usage, runtime, energy consumption, and FLOPs. The latter three values are collected using the Linux profiler tool \href{https://www.man7.org/linux/man-pages/man1/perf-stat.1.html}{\textit{perf}}. For quantifying memory usage, we use the Python package \href{https://docs.python.org/3/library/tracemalloc.html}{tracemalloc}. The correctness is evaluated by LeetCode and by human evaluation.

\paragraph{\textbf{Code Correctness}} Once we generate the initial and optimized solutions to the tasks using the three code generation models, we must validate the code's correctness. This can be done by a human analyzing the code manually or by using a set of test cases defining the expected output of an implemented function for a certain input. In our experiments, we used the automatic tests provided by LeetCode. Users on the LeetCode website can test their solution using a limited number of example inputs to promptly determine their code's correctness. These sample inputs typically only address general correctness. Once the user has performed a brief validation using the provided examples, they can submit their code. Subsequently, the code undergoes a comprehensive evaluation through tests encompassing critical inputs and boundary scenarios. To assess accuracy, we submit the code produced by each model (and for each prompt) to the LeetCode platform. Upon successful submission, we proceed with the examining the sustainability metrics.

\paragraph{\textbf{Runtime}}
We use the Linux profiler tool \textit{perf}~\cite{perf} to evaluate the generated code's runtime. Perf is a comprehensive performance analysis tool for applications, including both executables and terminal commands. It provides in-depth metrics such as the count of executed instructions, page faults, and branch misses. Additionally, it measures key parameters like execution runtime, energy usage, and Floating Point Operations per Second (FLOPs). These aspects will be explored in greater detail subsequently. Since the runtime is not a constant value for each execution, we monitor (through perf) and average the recorded runtime values. For this study, we chose to sample 10 runtime measurements.

\paragraph{\textbf{Energy Consumption}}
Energy consumption measurement falls into two types: hardware-based methods involving power meters and software-based methods using hardware APIs, which are more affordable~\cite{MANCEBO2021106560}. Here, we use again the \textit{perf} tool to measure the energy consumed through the code execution, similar to measuring the runtime. Regarding \textit{perf}'s configuration, we use the \texttt{power/energy-pkg/} event.
Similarly, to the runtime monitoring, we perform ten measurement sampling rounds. This approach yields the average energy consumption in Joules per code execution, reflecting the entire system's energy use. We also minimize the number of running background tasks that, in combination with the repeated code executions, can partially mitigate perf's measurement sensitivity.

\paragraph{\textbf{FLOPs}}
For evaluating FLOPs, we again configure the \textit{perf} tool with the \texttt{fp\_arith\_inst\_retired.scalar\_double} event to measure floating-point operations during code execution. Given that the number of floating-point operations remains constant for the same code and input, there is no need for repeated runs. 
\noindent\paragraph{\textbf{Memory Usage}}


To monitor the memory usage of the generated code execution, we use the Python package \textit{tracemalloc}, a built-in memory profiler allowing to trace memory allocations. In our evaluation, we use the \texttt{get\_traced\_memory()} function, which outputs the amount of currently allocated memory and the peak allocated memory. We start the memory tracing before invoking the functions related to the coding problem solution and output the peak allocated memory after the last related function call.

\begin{figure*}[t]
    \centering
    \includegraphics[width=.8\linewidth]{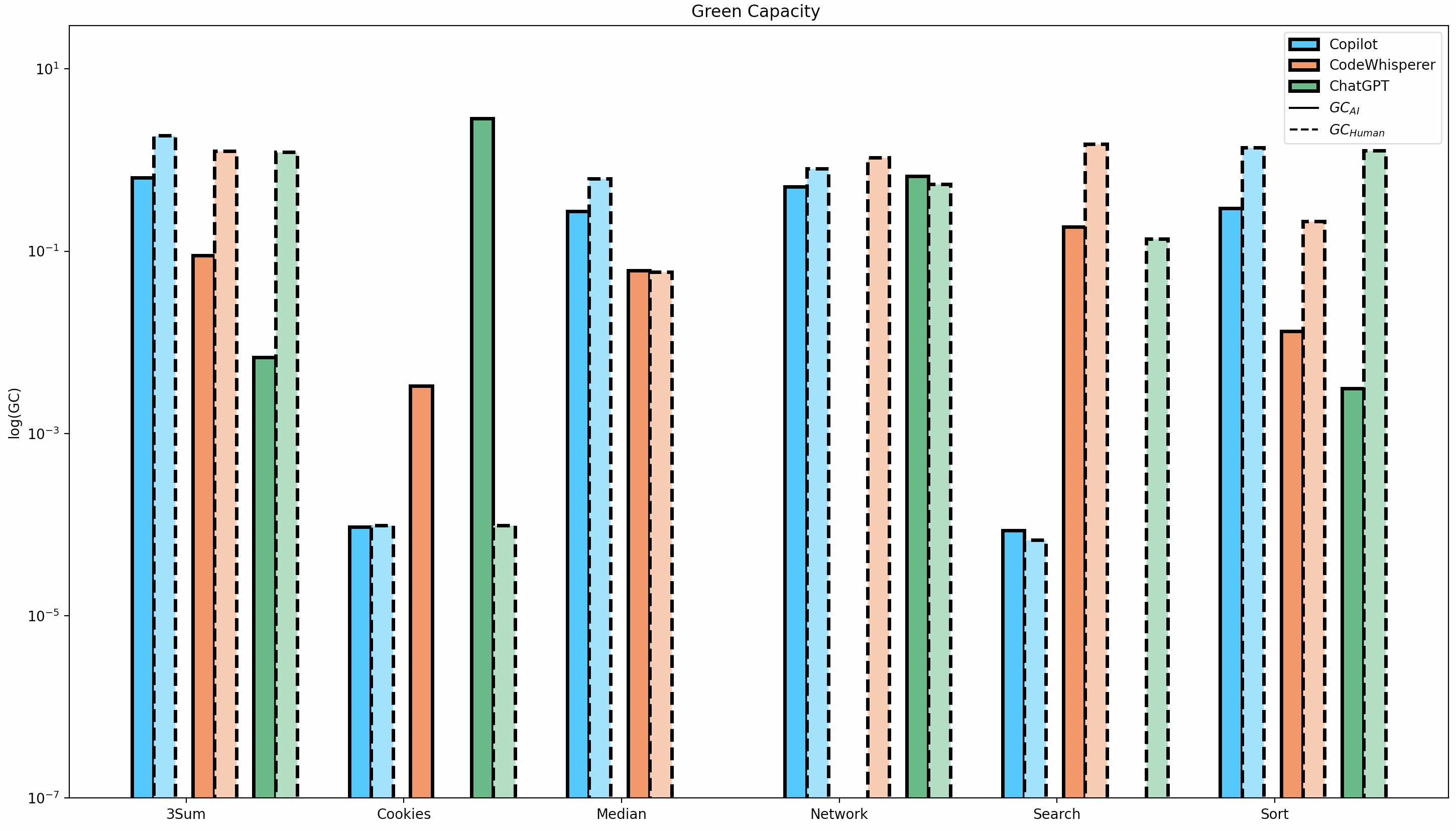}
    \caption{Comparison plot of Green Capacity values among the three automated code generation models (Copilot, CodeWhisperer, ChatGPT) throughout six coding problems (3Sum, Cookies, Median, Network, Search, Sort). For each reported $\text{GC}_\text{AI}$ value (depicted with the full line bar), we append on its right the computed $\text{GC}_\text{Human}$, that compares the human submission with the initially generated code for each model. A higher left value implies that the code generation model performed better than the human submission throughout the sustainability metrics. Absence of data denotes a value of 0.}
    \label{fig:hist_overall}
\end{figure*}

\section{Results $\&$ Discussion}

In this section, we analyze the evaluation results across the different models and coding problems. We first compare the Green Capacity values for each model based on their ability to consider the sustainability metrics in the code generation and the Green capacity of the submission (that is runtime-optimized) with respect to the initially auto-generated codes (one for each model).
We then analyze the PD values of the initial and the optimized AI-generated codes for specified sustainability metrics.

\subsection*{Comparative Analysis of Green Capacity}



Figure~\ref{fig:hist_overall} presents a comparative analysis of Green Capacity values observed across different code generation tools for each coding problem. It also contrasts the performance of these AI models with human submissions. Initially, it is evident that the three tools—ChatGPT, Copilot, and CodeWhisperer—demonstrate a partial understanding of the optimization criteria. Notably, ChatGPT and Copilot exhibit high Green Capacity scores in most problems. 

A positive Green Capacity value shows enhanced sustainability and a beneficial impact from requesting code optimized with respect to a given sustainability metric. In addition, the optimization techniques applied to ChatGPT significantly enhance its performance in the Cookies problem. This trend is also observed in most of the coding problems, where ChatGPT displays competitive results against human submissions. A second observation is that selecting the runtime-optimized human submissions leads to generally surpassing the AI models in performance. We note here that the human submissions are not the average submissions of a programmer but the best submitted solution in Leetcode and thus difficult to be surpassed. Specifically, Copilot's outputs are comparable to the best human submissions in 2 out of the five coding challenges.

Furthermore, it is noteworthy that although the human submissions are mainly focused on runtime optimization (i.e., achieving top runtime scores in LeetCode), they can surpass multiple times the optimized code instances. For instance, human submissions achieve higher GC values than Copilot in the majority of the coding problems. This implies that the \textit{same} human submission can achieve improvements over \textit{multiple} sustainability metrics (without explicitly optimizing them) over the initially generated implementation. These results indicate that there is a correlation among the sustainability metrics (i.e., higher values of one metric may be related to higher values of another one).

\begin{figure}[t]
    \centering
    \includegraphics[width=1\linewidth]{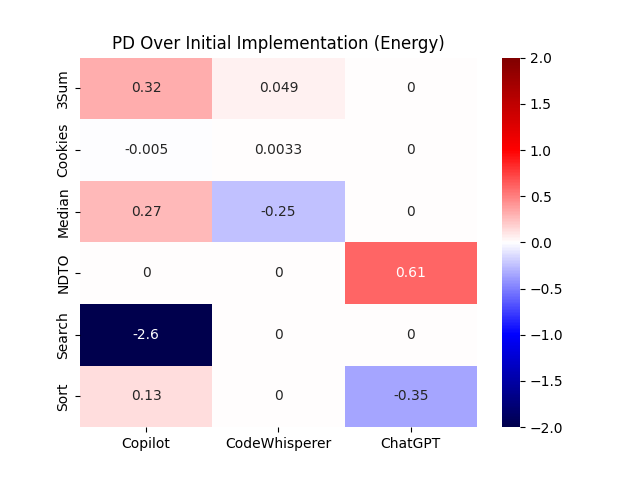}
    \caption{Heat Map Displaying Performance Delta (PD) in Energy Across Various Tasks and Tools: This map visually represents the difference between initial and optimized solutions. Each cell indicates the PD's contribution to Green Capacity (GC) values for a specific problem task and AI tool. Positive PD values enhance GC, signifying a beneficial impact on sustainability. In contrast, negative or zero PD values indicate a non-contributory effect on GC.}
    \label{fig:PD_Energy}
\end{figure}

\begin{figure}[t]
    \centering
    \includegraphics[width=1\linewidth]{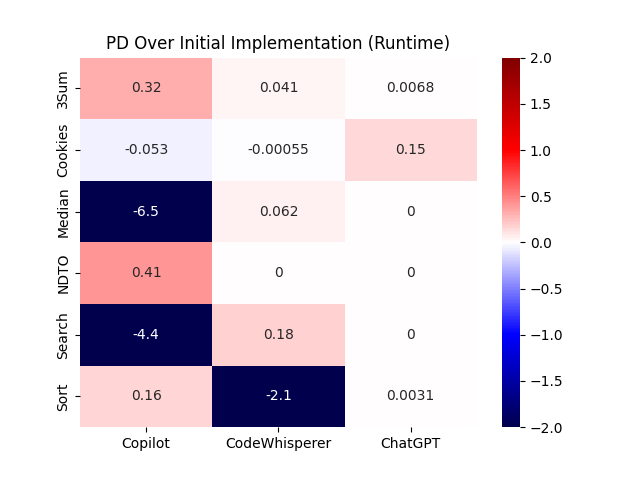}
    \caption{Heat Map Displaying Performance Delta (PD) in Runtime Across Various Tasks and Tools: This map illustrates the variance in energy efficiency across different AI tools and coding problems. It highlights instances where tools like ChatGPT and Copilot either fail to produce efficient code or generate code less efficient than the initial implementation, underscoring the challenges in aligning code optimization with the energy efficiency sustainability metric.}
    \label{fig:PD_Runtime}
\end{figure}

\begin{figure}[t]
    \centering
    \includegraphics[width=1\linewidth]{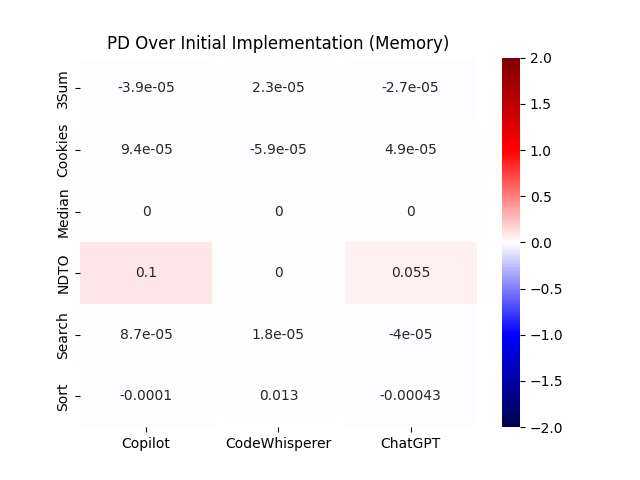}
    \caption{Heat Map Displaying Performance Delta (PD) in Memory Across Various Tasks and Tools: This map shows the difference between initial and optimized solutions. Each cell indicates the PD's contribution to GC values for a specific problem task and AI tool. Positive PD values enhance GC, signifying a beneficial impact on sustainability. We observe a high variance of non-positive PD values, showcasing the unawareness of the models to the given metric.}
    \label{fig:PD_Memory}
\end{figure}

\subsection*{Further investigation of optimization criteria}

While GC entails aggregated information over all optimization criteria, we can further investigate the behavior of the examined code-generation tools for each sustainability metric. We leverage the PD function values (see Equation~\ref{eq:pd}), that show the improvements (or lack thereof) achieved by requesting solutions optimized for a given sustainability metric. The heatmaps for energy consumption, runtime, and memory, are shown in Figures~\ref{fig:PD_Energy}, \ref{fig:PD_Runtime}, \ref{fig:PD_Memory}. 


The PD (Performance Delta) values provide additional insight on the code generation tools' capacity to integrate sustainability metrics. PD data includes both positive and negative values, where a positive PD value indicates that the optimized code outperforms the initially auto-generated counterpart on a given sustainability metric. For instance, as shown in Figure~\ref{fig:PD_Energy}, Copilot achieves a PD of 0.61 on the 'NDTO' coding problem. This implies that Copilot's optimized code consumes 0.61 times less energy than the initially generated code. Conversely, negative PD values highlight the code generation tool's lack of understanding of the impact of energy consumption on code efficiency for the given task.

Figure~\ref{fig:PD_Energy} shows the PD results with respect to Energy Consumption. We observe that in multiple cases, CodeWhisperer and ChatGPT fail to produce or optimize codes (for $3$ and $4$ coding problems, respectively) based on this sustainability metric. On the other hand, Copilot showcases the best behavior across the coding problems. Similarly, in Figure~\ref{fig:PD_Runtime} shows the PD values for to the runtime metric. ChatGPT fails to produce a valid code output for $3$ out of $6$ coding problems. Moreover, Copilot outputs code whose runtime is worse than the initial implementation. This variance across the coding problems shows that the code generation tool is not yet able to identify the impact of runtime as a sustainability metric for code optimization. Finally, Figure~\ref{fig:PD_Memory} depicts the PD values for the memory metric. We observe a high number of zero PD values, highlighting that memory was not a significant contributor to the Green capacity values described here. The importance of this sustainability metric and the ability of code generation tools to optimize it could be further investigated by performing analysis on different coding problems; however, this is outside the scope of our study.

\section{Conclusion}

With the growing demand for data services, addressing the environmental impact of software development has become an increasing area of interest. Large language models have the potential to be widely deployed, triggering a possibly growing carbon footprint. Green coding is often mentioned as a strategy to help reduce the environmental impact of software development, where sustainability is integrated into the code generation process. Here, we look into recently developed LLMs that have the ability to generate code and investigate whether these generative AI models can integrate sustainability in their generation of code in response to selected prompts. Sustainability is quantified using a newly defined metric of “green capacity”, a relative value that captures the ability of a code to reduce FLOPs, runtime, memory use, and energy consumption of successfully generated code. We compare the green capacity of a human to that of three LLMs – GitHub Copilot, OpenAI ChatGPT-3, and Amazon CodeWhisperer. We find that the AI models achieve increases in the sustainability of the code they generate when asked to optimize sustainability metrics. The AI tools produce a slightly computationally heavier and thus less environmentally conscious code than the optimized human solution chosen from Leetcode.


These findings highlight that LLMs do not yet have the sustainability awareness that would allow for reductions in carbon emissions from their use. Given the significant carbon footprint associated with training these models, emissions reductions in their application would be beneficial. We note that the models studied here are quite recent, allowing for progress in sustainability awareness that could be achieved by embedding sustainability goals in the training of models to learn to generate more sustainable outputs. Moreover, there is a continuous evolution in the automated code generation tools, such as new versions or new tools being developed, such as \href{https://www.ibm.com/products/watsonx-code-assistant}{IBM Watson X Code Assistant}, and \href{https://chat.openai.com/}{OpenAI ChatGPT 4.0}, which are going to be considered in future analysis. 

For future work, we aim to: a) expand our experiments by including a broader range of coding problems; b) enhance the prompt engineering methods for evaluation, acknowledging the code generation models' sensitivity to optimization prompts; c) broaden the range of sustainability metrics used to assess green capacity; and d) investigate the variability in factors like grid carbon intensity, PUE (Power Usage Effectiveness), or hardware specifications.

\section{Acknowledgements}
 

This work inspired a development project between Mercedes-Benz and TWT. We express our gratitude to IBM, Microsoft, and GitHub for the discussions on automated code generation that we had as part of this project.
We express our gratitude to Raphaela Löbel from TWT, Maria Stasinou from NIKI Ltd Digital Engineering, and Viet Dung Le from Mercedes-Benz for their engaging and informative discussions, which significantly influenced the project’s scope.

\nocite{*} 
\printbibliography



\end{document}